\documentclass[pra,onecolumn,floatfix,a4paper,superscriptaddress]{revtex4}
\usepackage{bm,color,graphicx,amsmath,txfonts}

\usepackage[colorlinks, citecolor=blue,linkcolor=blue]{hyperref}

\newcommand{\ic}{{i}}
\newcommand{\e}{{e}}

\begin{document}

\title{Quantum synchronization and entanglement of two magnon modes in a magnomechanical system}

\author{Hamza Harraf}
\affiliation{LPHE-Modeling and Simulation, Faculty of Sciences, Mohammed V University in Rabat, Rabat, Morocco.}	
\author{Mohamed Amazioug} \thanks{amazioug@gmail.com}
\affiliation{LPTHE-Department of Physics, Faculty of Sciences, Ibnou Zohr University, Agadir 80000, Morocco}
\author{Rachid Ahl Laamara}
\affiliation{LPHE-Modeling and Simulation, Faculty of Sciences, Mohammed V University in Rabat, Rabat, Morocco}
\affiliation{Centre of Physics and Mathematics, CPM, Faculty of Sciences, Mohammed V University in Rabat, Rabat, Morocco}

\begin{abstract}

We investigate theoretically the improvement of the entanglement between the indirectly coupled two magnon modes in a magnomechanical system with magnon squeezing. We quantify the degree of entanglement via logarithmic negativity between two magnon modes. We show a significant enhancement of entanglement via magnon squeezing. Additionally, the entanglement of two magnons decreases monotonically under thermal effects. We demonstrate that with an increasing photon tunneling rate, entanglement is robust and resistant to thermal effects. We use purity as a witness to the mixing between the two magnon modes. We show that synchronization and purity are very robust against thermal effects rather than entanglement. We examine the relationship between quantum entanglement, purity, and quantum synchronization in both steady and dynamic states. According to our results, this scheme could be a promising platform for studying macroscopic quantum phenomena.


\end{abstract}

\date{\today}

\maketitle

\section{Introduction}

Huygens made the initial prediction about the synchronization phenomenon in 1665 \cite{CHuygens1665} by studying the synchronized motion between two pendulum clocks with a common support. The concept of synchronization finds its application in a diverse range of fields including physical, biological, chemical, and social systems \cite{SHStrogatz2003,SBregni2002,ERanta1999,TWomelsdrof2007,JBuck1968,MToiya2010}. Recently, there has been a special focus on the phenomenon of spontaneous synchronization in the quantum realm. This is connected to the presence of quantum correlations in multiparite systems \cite{AMari2013}. In this way, synchronization and entanglement, quantum discord, and mutual information have all been studied in \cite{FGalve2017}. Several intriguing technological applications of spontaneous synchronization exist, such as high precision clocks \cite{DAntonio2012}, sensing \cite{IBargatin2012}, information processing \cite{KMakino2016}, quantum communications \cite{MMorelli2007}, cavity quantum electrodynamics \cite{OVZhirovand2009,VAmeri2015}, atomic ensembles \cite{MXu2014,MXuand2015,MRHush2015}, Bose Einstein condensation \cite{MSamoylova2015}, superconducting circuit systems \cite{YGul2016,FQuijandria2013} and encrypted communications \cite{LMPecora1990}. Recently, Amari et \textit{al} quantified the quantum synchronization for continuous variables \cite{AMari2013}.\\

Quantum correlations play a crucial role in quantum information processing and quantum communication. Entanglement is one of paramount importance for advanced  information processing such as quantum teleportation \cite{CHBennett1993}, superdense coding \cite{CHBennett2005}, telecloning \cite{VScarani2005} and quantum cryptography \cite{AKEkert1991}. The amount of entanglement between bipartite Gaussian mixed states is measured using logarithmic negativity \cite{GVidal2002,GAdesso2004}. Quantum optomechanical systems essentially require interaction between optical and mechanical modes through radiation pressure. This interaction is crucial for the creation of entangled states and the manipulation of quantum information. The ability to control and measure these systems has led to advancements in precision measurement, quantum computing, and quantum sensing. Optomechanical systems have also been used to study fundamental questions in physics, such as the nature of quantum measurement and the limits of classical mechanics. In recent years, researchers have explored new ways to enhance optomechanical interaction through novel materials and geometries. These developments have opened up new avenues for exploring the quantum world and may lead to breakthroughs in fields ranging from telecommunications to fundamental physics. As our understanding of these systems continues to grow, we can expect even more exciting discoveries in the years ahead. Furthermore, it is frequently claimed that the decoherence phenomenon that arises when a quantum system interacts with its environment is the main obstacle to maintaining entanglement \cite{WHZurek2003}. The best platform for studying quantum spontaneous synchronization involving continuous variables is provided by optomechanical systems \cite{APikovsky2003,SHStrogatz2001,CGLiao2019}.\\

Recently, the magnon, which are the quanta of collective spin excitations in yttrium iron garnet $(\text{Y}_3\text{Fe}_5\text{O}_{12}; \text{YIG})$, are of utmost significance because of their high spin density, low damping rate, and excellent tunability. They also have potential applications in spintronics and quantum computing. This makes yttrium iron garnet a promising candidate for the next generation of spin-based devices. Furthermore, yttrium iron garnet is a non-toxic and biocompatible material, making it suitable for biomedical applications such as magnetic resonance. Bloch was the first to describe the magnon \cite{Bloch1930}, which is the collective spin wave excitation carrying quantized energy in a magnetically ordered ground state. Besides, cavity magnetoelectronics has attracted considerable attention recently because it provides a strong platform on which a ferrimagnetic crystal YIG sphere is coupled with a microwave cavity \cite{DWWang2023,MSDing2022,DLachanceQuirion2019,HYYan2022}, phonons, \cite{JLie2018} and photons \cite{AOsada2018,AOsada2016,XFZhang2016}. By the magnetostrictive force, a magnon mode (spin wave) is combined with a vibratory deformation mode of a ferromagnet (or ferrimagnet) and a microwave cavity mode by the interaction of magnetic dipoles in cavity magnetomechanics. This allows for the conversion of microwave photons to magnons and vice versa, enabling the development of hybrid quantum systems. For a large ferromagnet, the magnetostrictive interaction is a dispersion interaction similar to radiation pressure, where the mechanical mode frequency is much lower than the magnon frequency \cite{ZYFan2022,XZhang2016}. Recently, quantum correlations with the cavity magnonics is of paramount importance in quantum information processing \cite{Li2018,Li2019,Nair2020,Zhou2022,Asjad2023,Chabar24,Amazioug25,Mathkoor2025}.
\\

In this work, we study theoretically the enhancement of entanglement between the two indirectly coupled magnon modes in a cavity magnomechanics with squeezed magnon. We shall use the logarithmic negativity as a witness of nonclassical correlations of two-mode CV Gaussian mixed states in both steady and dynamical regime. We shall employ the purity to quantify the mixdness between the two magnon modes. We shall quantify a quantum complete synchronization and quantum-phase synchronization between the two magnon modes. We shall show the significant enhancement of the continuous variable entanglement between the two magnon mode via the squeezing of the magnon mode and photon tunneling. The entanglement will be robust under thermal effect by increasing the squeezing parameter $\lambda$. The relationship between entanglement, purity, and synchronization will be discussed. \\

The paper is organized as follows. In Sect. 2, we give the Model and dynamical equations of magnomechanical system under consideration. We study the enhancement of entanglement and the relationship between entanglement, purity, and synchronization under some system parameters in Sect. 3. The results and discussions are given in Results and discussion section. Concluding remarks close this paper.

\begin{figure}[t]\label{fig0}
\hskip-1.0cm\includegraphics[width=0.7\linewidth]{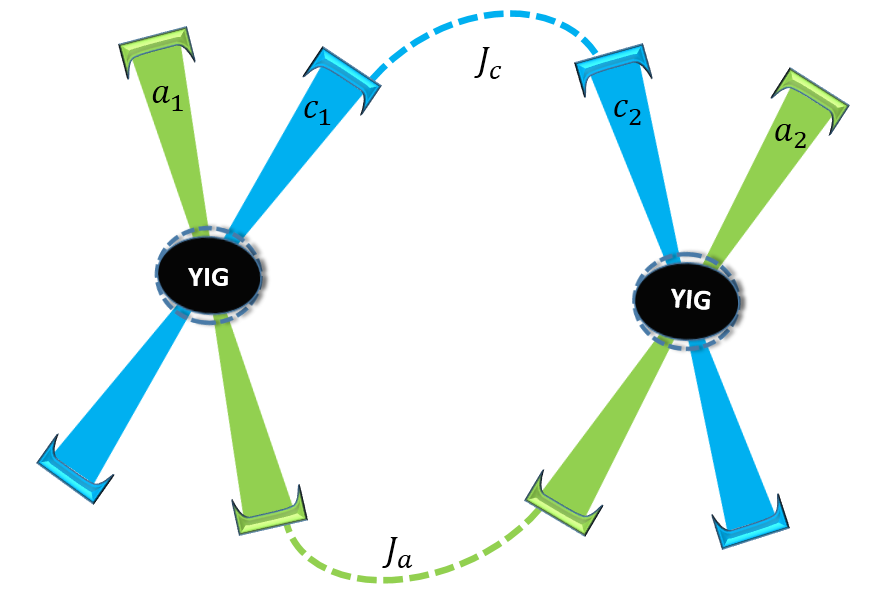} 
\caption{Schematics of hybrid cavity-magnon array with two sites and two YIG spheres is used, where each site has a magnon in a YIG sphere connected to two MW fields through a magnetic dipole interaction. Each sphere is positioned in a uniform bias magnetic field, close to the cavity mode maximum magnetic field, and is driven directly by a strong MW field (not shown) of amplitude $\Omega_j$ $(j=1,2)$ to improve magnon–phonon coupling. Also, in the left site we drive each cavity by coherent light (not shown) with amplitude $\mathcal{E}$. The cavity modes of adjacent sites become coupled to each other through controllable rates of tunneling. The rate at which particles jump between neighboring cavity $a$ is denoted as $J_a$, whereas the rate at which particles hop from cavity $c_1$ to $c_{2}$ is denoted as $J_c$.}
\end{figure}

\section{The Model and dynamical equation}\label{Model}

We consider a hybrid cavity-magnon array with two sites, each with one magnon that couples to two microwave (MW) fields via a magnetic dipole interaction, which can be very strong \cite{Strong1,Strong2,Strong3,Strong4,Strong5,Strong6,Tobar2,Tobar3}. We regard the YIG spheres as much smaller than microwave wavelengths, which allows us to safely ignore the influence of radiation pressure. In uniform bias magnetic fields, the magnon modes in YIG spheres are squeezing and approaching the maximum magnetic fields of the cavity modes, which excite the magnon modes in the spheres and couple them to the cavity modes. The two sites are coupled via photon tunneling as adopted in Fig. \ref{fig0}. The Hamiltonian of the system in a rotating frame at frequency $\omega_j$ is written as
\begin{eqnarray} \label{Hamilt}
\mathcal{H} =\underset{j=1,2}{\sum }\Big\{\Delta _{a_{j}}a_{j}^{\dag }a_{j} + \Delta _{c_{j}}c_{j}^{\dag }c_{j}+\Delta _{d_{j}}d_{j}^{\dag }d_{j}+\Delta _{b_{j}}b_{j}^{\dag }b_{j}+\frac{\omega_{b_j}}{2}(q_j^2+p_j^2)+g_{db_j} d_j^+d_j q_j+g_{cd_j} \big( c_{j}d_j^{\dag }+c_{j}^{\dag }d_j\big)+\nonumber\\
 g_{ad_j} \big( a_{j}d_j^{\dag }+a_{j}^{\dag }d_j\big)+\ic \lambda_j (\e^{\ic \theta}d_j^{+2}-\e^{-\ic \theta}d_j^{2})+\Omega_{j}(d^\dag_{j}-d_{j})\Big\}+ \mathcal{E}(a^\dag_{1}+a_{1})+\mathcal{E}(c^\dag_{1}+c_{1}) - J_a (a_1^+a_2 + a_2^+a_1)- J_c (c_1^+c_2 + c_2^+c_1),
\end{eqnarray}
where $\Delta _{a(c)_{j}}=\omega_{a(c)_{j}}-\omega _{_{j}},\Delta _{d_{j}}=\omega _{d_{j}}-\omega _{_{j}}$. $c_{j}$ ($c_{j}^{\dag }$) and $d_{j}$ ($d_{j}^{\dag }$) are, respectively, annihilation (creation) operators for the $jth$ cavity and $jth$ magnon modes, with $\big[\mathbf{O}, \mathbf{O}^+\big] =1$ ($\mathbf{O}\,{=}\,a_j,c_j, d_j$). $\omega_j$ is the frequency of the $jth$ mode of the input field. The resonance frequencies of the jth cavity mode (magnon mode) are denoted by $\omega_{c_{j}}$ ($\omega_{d_{j}}$). The external bias magnetic field $H_j$ and the gyromagnetic ratio $\beta$ throught $\omega _{d_{j}}=\beta H_j$ determine the frequency of the magnon mode $\omega_{d_{j}}$. $g_{ad_j}$ and $g_{cd_j}$ represent the rate of coupling between the jth cavity and jth magnon modes. $\Omega_j =\frac{\sqrt{5}}{4} \gamma_j \! \sqrt{N_j} B_{0 _j}$ is the Rabi frequency describes the coupling strength of the drive magnetic field (with $B_{0_j}$ and $\omega_{0_j}$ are respectively the amplitude and frequency) with the magnon mode, where $\gamma_j/2\pi= 28$ GHz/T, and the total number of spins $N_j=\rho V_j$ with $V_j$ the volume of the $jth$ sphere and $\rho=4.22 \times 10^{27}$ m$^{-3}$ the spin density of the YIG. According to the low-lying excitation hypothesis, $\langle d_j^{\dag} d_j \rangle \ll 2N_j s$, where $s=\frac{5}{2}$ is the spin number of the ground state Fe$^{3+}$ ion in YIG, the Rabi frequency $\Omega_j$ is derived. We have the squeezing parameter $\lambda_j$ and the phase $\theta$ for the $jth$ magnon mode squeezing. The magnon squeezing can be achieved by transferring squeezing from a squeezed-vacuum microwave field \cite{JLi2019}, or by the intrinsic nonlinearity of the magnetostriction (the so-called ponderomotive-like squeezing) \cite{JLiarxiv}, or by the anisotropy of the ferromagnet \cite{HYYuanarxiv,AKamara2016}, etc. The QLEs of the system are given by
\begin{eqnarray} \label{nonl} 
\dot{a_1}&=&-(i \Delta_{a_1}+\kappa_{a_1}) a_{1}-ig_{a_1}d_1-i\mathcal{E}_{1}+iJ_a a_2+\sqrt{2\kappa_{a_1}}a_1^{in},\nonumber\\
\dot{a_2}&=&-(i \Delta_{a_2}+\kappa_{a_2}) a_{2}-ig_{a_2}d_2+iJ_a a_1+\sqrt{2\kappa_{a_2}}a_2^{in},\nonumber\\
\dot{c}_{1}&=&-(i\Delta_{c_1}+\kappa_{c_1})c_1-ig_{c_1}d_1-i\mathcal{E}_{1}+iJ_a a_2+\sqrt{2\kappa_{c_1}}c_{1}^{in},\nonumber\\
\dot{c}_{2}&=&-(i\Delta_{c_2}+\kappa_{c_2})c_2-ig_{c_2}d_2+iJ_a a_1+\sqrt{2\kappa_{c_2}}c_{2}^{in},\nonumber\\
\dot{m}_j&=&-(i\Delta_{d_j}+\kappa_{d_j})d_j-ig_{a_j}a_j-ig_{c_j}c_j+2\lambda_j \e^{\ic \theta} d^+_j +\Omega_j+\sqrt{2\kappa_{d_j}}d_j^{in}, \nonumber\\
\dot{q}_j&=& \omega_{b_j} p_j,  \nonumber  \\
\dot{p}_j&=& - \omega_{b_j} q_j - \gamma_{b_j} p_j - g_{db_j} d_j^{\dag}d_j + \chi_j, 
\end{eqnarray}
where ($a_{j}^{in}$, $c_{j}^{in}$, $d_{j}^{in}$) are the input noise operators for the $jth$ cavity mode and $jth$ magnon mode, respectively, and $\kappa _{a(c)_j}$ ($\kappa _{d_{j}}$) are the decay rates of the $jth$ cavity mode and $jth$ magnon mode, respectively. In the time domain, the two cavity input noise operators $x_j^{in}$ ($x=a,c$) have the following non-zero correlations: $x^{in}_{j}$ and $x^{in+}_{j}$, which are given by \cite{JLi2020} 
\begin{equation} \label{eq:6} 
\langle x^{in}_{j}(t)x^{in+}_{j}(t'); x^{in+}_{j}(t)x^{in}_{j}(t')\rangle = (N_{x_{j}}+1; N_{x_{j}})\delta(t-t')
\end{equation}
The non-zero correlations of the magnon input noise operators $d_j^{in}$ are as follows
\begin{equation} \label{eq:4} 
\langle d_{j}^{in}(t) d_{j}^{in+}(t'); d_{j}^{in+}(t)d_{j}^{in}(t') \rangle = (N_{m_{j}}+1; N_{m_{j}})\delta(t-t')
\end{equation}
where $N_{x(d)_{j}}=\left[\exp \left(\frac{\hbar \omega _{x(d)_{j}}}{k_{B}T}\right)-1\right]^{-1}$ is the $jth$ mode's equilibrium mean thermal magnon number, $T$ is the environmental temperature, and $k_B$ is the Boltzmann constant.
\begin{equation} \label{eq:4} 
\langle \chi_{j}(t)\chi_{j}(t')+ \chi_{j}(t')\chi_{j}(t) \rangle = \gamma_{b_j}(2n_{b_{j}}+1)\delta(t-t').
\end{equation}

The nonlinear quantum Langevin equations in Eq.(\ref{nonl}) can be linearized by considering the operators as the sum of the expectation plus quantum fluctuations. In order to linearize these equations, one can write the annihilation operators as $ \mathbf{O} = \langle \mathbf{O} \rangle + \delta \mathbf{O} $
\begin{eqnarray} \label{eq:4} 
-(i \Delta_{a_1}+\kappa_{a_1}) \langle a_{1} \rangle -ig_{a_1}\langle d_{1} \rangle-i\mathcal{E}_{1}+iJ_a \langle a_{2} \rangle = 0  \nonumber \\
-(i \Delta_{a_2}+\kappa_{a_2}) \langle a_{2} \rangle -ig_{a_2}\langle d_{2} \rangle+iJ_a \langle a_{1} \rangle = 0  \nonumber \\
-(i \Delta_{c_1}+\kappa_{c_1}) \langle c_{1} \rangle -ig_{c_1}\langle d_{1} \rangle-i\mathcal{E}_{1}+iJ_c \langle c_{2} \rangle = 0  \nonumber \\
-(i \Delta_{c_2}+\kappa_{c_2}) \langle c_{2} \rangle -ig_{c_2}\langle d_{2} \rangle+iJ_c \langle c_{1} \rangle = 0  \nonumber \\
-(i\Delta_{d_j}+\kappa_{d_j})\langle d_{j} \rangle-ig_{a_j}\langle a_{j} \rangle-ig_{c_j}\langle c_{j} \rangle+2\lambda_j \e^{\ic \theta} \langle d_{j} \rangle^\star +\Omega_j =0 \nonumber  \\
- \omega_{b_j} \langle q_{j} \rangle - g_{db_j} |\langle d_{j} \rangle|^2 = 0
\end{eqnarray}
The linearized QLEs corresponding to the dynamics of quantum fluctuations are given by
\begin{eqnarray} \label{LQLEs}
\delta\dot{a_j}&=&-(i \Delta_{a_j}+\kappa_{a_j}) \delta a_{j}-ig_{a_j}\delta d_j+iJ_a \delta a_s+\sqrt{2\kappa_{a_j}}\delta a_j^{in},\quad j\neq s \nonumber\\
\delta\dot{c}_{j}&=&-(i\Delta_{c_j}+\kappa_{c_j})\delta c_j-ig_{c_j}\delta d_j+iJ_c \delta c_s+\sqrt{2\kappa_{c_j}}\delta c_{j}^{in},\quad j\neq s \nonumber\\
\delta\dot{d}_j&=&-(i\Delta_{d_j}+\kappa_{d_j})\delta d_j-ig_{a_j}\delta a_j-ig_{c_j}\delta c_j+2\lambda_j \e^{\ic \theta} \delta d^+_j +\sqrt{2\kappa_{d_j}}\delta d_j^{in}, \nonumber\\
\delta\dot{q}_j&=& \omega_{b_j} \delta p_j,  \nonumber  \\
\delta\dot{p}_j&=& - \omega_{b_j} \delta q_j - \gamma_{b_j} \delta p_j + iG_{db_j} (\delta d_j^{\dag} - \delta d_j)/\sqrt{2} + \chi_j, 
\end{eqnarray}
where $G_{db_j} = i\sqrt{2}g_{db_j}\langle d_{j} \rangle$. The linearized QLEs describing the quadrature fluctuations  
$(\delta X_{d_1},\delta P_{d_1},\delta X_{d_2},\delta P_{d_2},\delta q_1,\delta p_1, \delta q_2,\delta p_2, \delta X_{a_1},\delta P_{a_1},\delta X_{a_2},\delta P_{a_2},\delta X_{c_1},\delta P_{c_1},\delta X_{c_2},\delta P_{c_2})$,
with  $\delta X_{d_j}=(\delta d_{j}+\delta d_{j}^{\dag })/\sqrt{2},\delta P_{d_j}=i(\delta d_{j}^{\dag }-\delta d_{j})/\sqrt{2},\delta X_{a_j}=(\delta a_{j}+\delta a_{j}^{\dag })/\sqrt{2},\delta P_{a_j}=i(\delta a_{j}^{\dag }-\delta a_{j})/\sqrt{2}, \delta X_{c_j}=(\delta c_{j}+\delta c_{j}^{\dag })/\sqrt{2},\delta P_{c_j}=i(\delta c_{j}^{\dag }-\delta c_{j})/\sqrt{2},$ (similar definition for input noises $\delta X_{d_j}^{in}, \delta P_{d_j}^{in}$, $\delta X_{a_j}^{in}, \delta P_{a_j}^{in}$ and $\delta X_{c_j}^{in}, \delta P_{c_j}^{in}$), given by
\begin{equation}
\dot{r}(t)=\mathcal{K} r(t)+\zeta (t),  \label{MatrixForm}
\end{equation}
where $r(t)=[\delta X_{d_1},\delta P_{d_1},\delta X_{d_2},\delta P_{d_2},\delta q_{1,}\delta p_{1},\delta q_{2},\delta p_{2},\delta X_{a_1},\delta P_{a_1},\delta X_{a_2},\delta P_{a_2},\delta X_{c_1},\delta P_{c_1},\delta X_{c_2},\delta P_{c_2}]^{T}$, 

 $\zeta(t)=[\sqrt{2\kappa_{d_{1}}}X_{d_1}^{in},\sqrt{2\kappa_{d_{1}}}P_{d_1}^{in}, \sqrt{2\kappa_{d_{2}}}X_{d_2}^{in},\sqrt{2\kappa_{d_{2}}}P_{d_2}^{in},0,\chi,0,\chi ,\sqrt{2\kappa_{a_{1}}}X_{a_1}^{in},\sqrt{2\kappa_{a_{1}}}P_{a_1}^{in}, \sqrt{2\kappa_{a_{2}}}X_{a_2}^{in},\sqrt{2\kappa_{a_{2}}}P_{a_2}^{in},\\\sqrt{2\kappa_{c_{1}}}X_{c_1}^{in},\sqrt{2\kappa_{c_{1}}}P_{c_1}^{in}, \sqrt{2\kappa_{c_{2}}}X_{c_2}^{in},\sqrt{2\kappa_{c_{2}}}P_{c_2}^{in}]^{T}$, and 
$\mathcal{K}$ is a drift matrix, and its elements can obtained from equation \ref{LQLEs}. When the eigenvalues of the drift matrix $\mathcal{K}$ have negative real parts, the system is stable, according to the Routh-Hurwitz criterion \cite{EXDeJesus1987}. 

\section{Entanglement, Purity and synchronization}

The steady and dynamical state of the system, is completely described by an $16\times 16$ covariance matrix (CM). According to the Lyapunov equation we can derived the covariance matrix $\mathcal{C}$ of the system~\cite{DVitali2007} 
\begin{equation}
\mathcal{K} \mathcal{C} + \mathcal{C}\mathcal{K}^T = \dot{\mathcal{C}}-\cal{L},  \label{LyapEq}
\end{equation}
where $\mathcal{C}_{ik}(t)=\langle r_{i}(t)r_{k}(t^{\prime })+r_{k}(t^{\prime })r_{i}(t) \rangle/2$ and $\mathcal{U}$ is the diffuse matrix defined by $\mathcal{L}_{ik}\delta (t-t^{\prime})=\langle \zeta_{i}(t) \zeta_{k}(t^{\prime })+\zeta_{k}(t^{\prime })\zeta_{i}(t) \rangle/2$, given by $\mathcal{L}=\mathcal{L}_{db}\oplus \mathcal{L}_{ac}$,
where $\mathcal{L}_{db} = diag[\kappa_{d_1}(1+2N_{d_1}),\kappa_{d_1}(1+2N_{d_1}),\kappa_{d_2}(1+2N_{d_2}), \kappa_{d_2}(1+2N_{d_2}),0,\gamma_{b_1}(1+2n_{b_1}),0,\gamma_{b_2}(1+2n_{b_2})]$
and 
$\mathcal{L}_{ac} = diag[\kappa_{a_1}(1+2N_{a_1}),\kappa_{a_1}(1+2N_{a_1}),\kappa_{a_2}(1+2N_{a_2}), \kappa_{a_2}(1+2N_{a_2}),\kappa_{c_1}(1+2N_{c_1}),\kappa_{c_1}(1+2N_{c_1}),\kappa_{c_2}(1+2N_{c_2}), \kappa_{c_2}(1+2N_{c_2})]$.
The system's initial state is considered to be in the vacuum state. The covariance matrix for the two magnon modes is given by
\begin{equation} \label{eq:Sigmamm}
\mathcal{C}_{(dd)}=
\begin{pmatrix}
	\mathcal{X} & \mathcal{Z}  \\
    \mathcal{Z}^T & \mathcal{Y}  
\end{pmatrix}
\end{equation} 
The $2\times 2$ sub-matrices $\mathcal{X}$ and $\mathcal{Y}$ in Eq. (\ref{eq:Sigmamm}) characterize the autocorrelations of the two magnon modes, and the $2\times 2$ sub-matrix $\mathcal{Z}$ describes the cross-correlations of the two magnon modes. The logarithmic negativity $E_{dd}$ between the two magnon modes, it's written as \cite{GAdesso2004, Plenio}
\begin{equation} \label{eq:37}
	E_{dd}=\max[0,-\log(2\mu^-)]
\end{equation}
where $\mu^-$ represents the smallest symplectic eigenvalue of the partially transposed covariance matrix $\mathcal{C}_{(dd)}$ of two magnon modes 
\begin{equation} \label{eq:38}
\mu^-= \sqrt{\frac{\Gamma-\sqrt{\Gamma^2-4\det\mathcal{C}_{(dd)}}}{2}}   
\end{equation}
where the symbol $\Gamma$ is written as $\Gamma=\det \mathcal{X}+\det \mathcal{Y}-2\det \mathcal{Z}$. The two magnon modes are separable if $E_{dd}=0$.
The purity of a two magnon modes is denoted as
\begin{equation} \label{eq:38}
P = \frac{1}{4\sqrt{\det \mathcal{C}_{dd}}}   
\end{equation}
The quantum complete synchronization proposed by Mari et al. \cite{AMari2013} for two magnon modes is given by
\begin{equation} \label{eq:23}
S_c (t)=\frac{1}{\langle \delta X_-(t) ^2 + \delta P_-(t) ^2\rangle},
\end{equation}
where $\delta X_- (t)=[\delta X_{d_1} (t)-\delta X_{d_2} (t)]/\sqrt{2}$ and $\delta P_- (t)=[\delta P_{d_1} (t)-\delta P_{d_2} (t)]/\sqrt{2}$ are the error operators. If $S_c (t) = 0$ (fully unsynchronized) and if $S_c (t) = 1$ (fully synchronized). The quantum-phase synchronization of two magnon modes is defined as
\begin{equation} \label{eq:38}
S_p (t) = \frac{1}{2}\langle \delta P_- (t)^2 \rangle^{-1}  
\end{equation}
with $\delta P_- = [\delta P_{d_1} - \delta P_{d_2}]/\sqrt{2}$ characterize the difference between the two momentum operators. If $0 < S_p(t) \leq 1$, the phase synchronization is realized, if $ S_p(t)>1$ explain that the fluctuation is squeezed.

\section{Resultats and Discusion}

In this section, we will look at the steady and dynamical state quantum correlations, purity and synchronizations of two magnon modes under different effects using the experimental values reported in \cite{JLi2020}: $\omega_{b}/2\pi= 10$ MHz, $\omega_{a(c)}/2\pi= 10$ GHz, $\kappa_{a(c)}/2\pi= 1$ MHz, $\gamma_d/2\pi=100$ Hz and $g_{a(c)}/2\pi= 4.8$ MHz, for simplicity, we consider that the two sites are symmetric and $\theta=0$. 

According to \cite{JLi2021}, we use $250 \mu$m such as diameter of YIG sphere to obtain $N=3.5\times 10^{16}$. We consider $G_{db}/2\pi= 0.1$ MHz, and $\kappa_{d}=0.6\kappa_{c}$ implies the drive power $P \approx 0.45$ mW. This corresponding to the drive magnetic field $B_0 \approx 3.9 \times 10^{-5}$ T for $\Omega = 7.2\times 10^{14}$ Hz.

\begin{figure*}[htb]
\includegraphics[width=\textwidth]{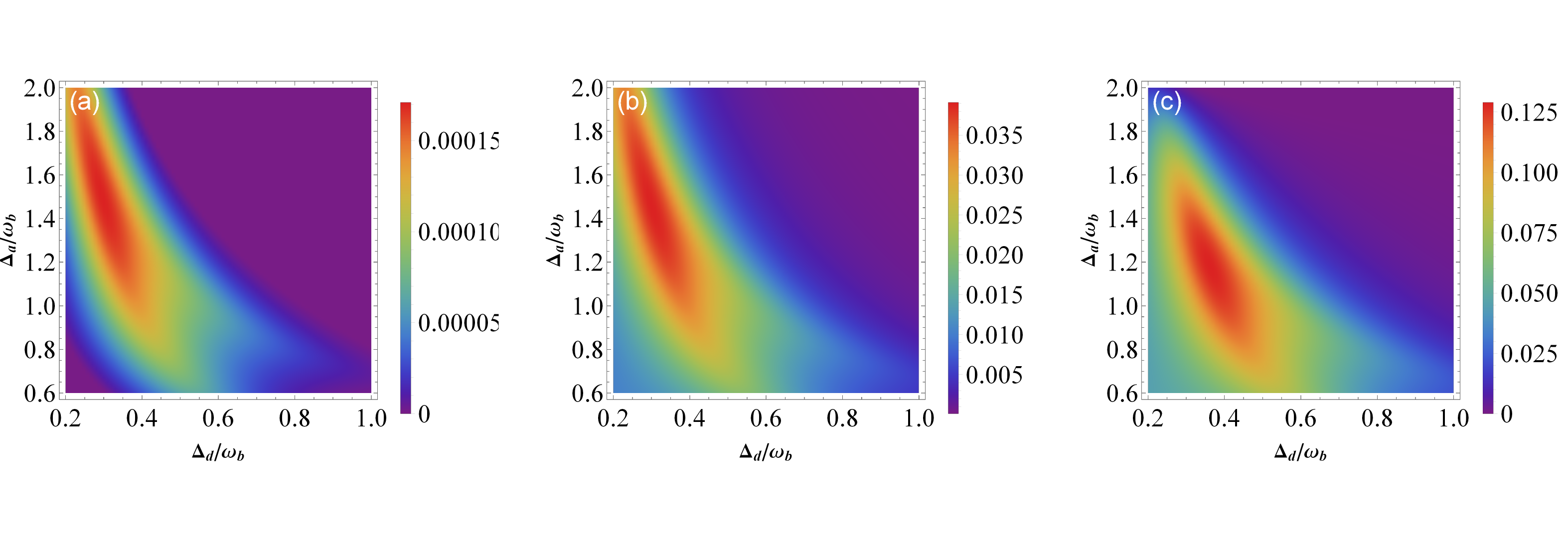}
\caption{Density plot of entanglement $E_{dd}$ between two magnon modes as a function of the detunings $\Delta_a$ and $\Delta_d$ with $J = 0.5g_a$ and $T=0.1$ mK. $\lambda=0$ in (a), $\lambda=0.005g_a$ in (b) and $\lambda=0.05g_a$ in (c).}
\label{fig1}
\end{figure*}

In Fig. \ref{fig1}, we have plotted the logarithmic negativity $E_{dd}$ of subsystem magnon-magnon versus the $\Delta_a=\Delta_c$ and $\Delta_d$ with and without magnon squeezing. We note first that magnon squeezing influence on the logarithmic negativity $E_{dd}$ value. We remark that when $\Delta_a=\omega_b$ and $\Delta_d=0.4\omega_b$, the entanglement between the two magnon modes is optimal as depicted in Fig. \ref{fig1} (b). Moreover, when $\Delta_a=\omega_b$ and $\Delta_d=0.4\omega_b$ the two magnon modes are separated. Thus, one can say that the presence of squeezing generate and is helpful to enhance the entanglement of two separate magnon modes. Compared with the case of without the squeezing ($\lambda =0$), the entanglement $E_{dd}$ can be significantly improved, and it can be seen that the maximum of the entanglement with the squeezing increases by 500 $\%$. 

\begin{figure*}[!htb]
\includegraphics[width=\linewidth]{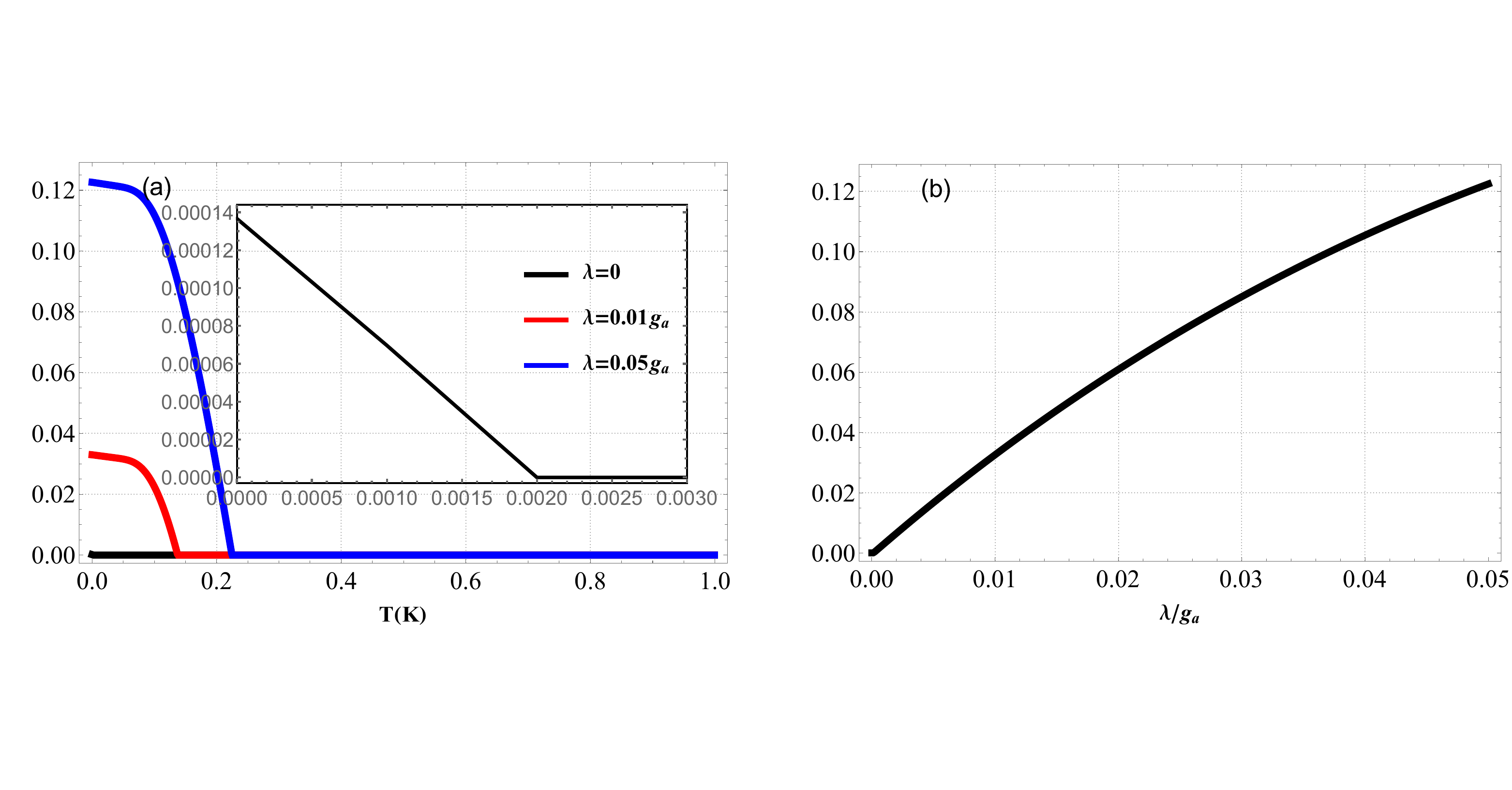}
\caption{(a) Plot of the logarithmic negativity $E_{dd}$ between the two magnon modes versus the $T$ for various value of $\lambda$ with $\Delta_a=\omega_b$ and $\Delta_d=0.4\omega_b$ and $J = 0.5g_a$. (b) Plot of the logarithmic negativity $E_{dd}$ between the two magnon modes vs $\lambda/g_a$ with $T = 0.1$ mK }
\label{fig2}
\end{figure*}

We plot in Fig. \ref{fig2}(a) the logarithmic negativity $E_{dd}$ between the two magnon modes as a functions of the temperature $T$ for various value of $\lambda$. The entanglement is diminishes under thermal effects via decoherent phenomenon. And the strong entanglement only exist at cryogenic temperature. Besides, quantum correlations are robust against thermal effects and survive for a wide range of temperature in the presence of squeezing. We observe in Fig. \ref{fig2}(b) a significant enhancement of the entanglement $E_{dd}$ with increasing $\lambda/g_a$. This can be explain the gradual increase of $\lambda$ from 0 means that the nonlinearity of the system is enhanced, resulting in the increase of entanglement. And when $\lambda=0$ (the absence of squeezing) the entanglement $E_{dd}=0$. Compared with the case of with the squeezing (the presence of squeezing) one can say that squeezing plays a crucial role in generating the entanglement between the two magnon. 

\begin{figure*}[!htb]
\includegraphics[width=0.7\linewidth]{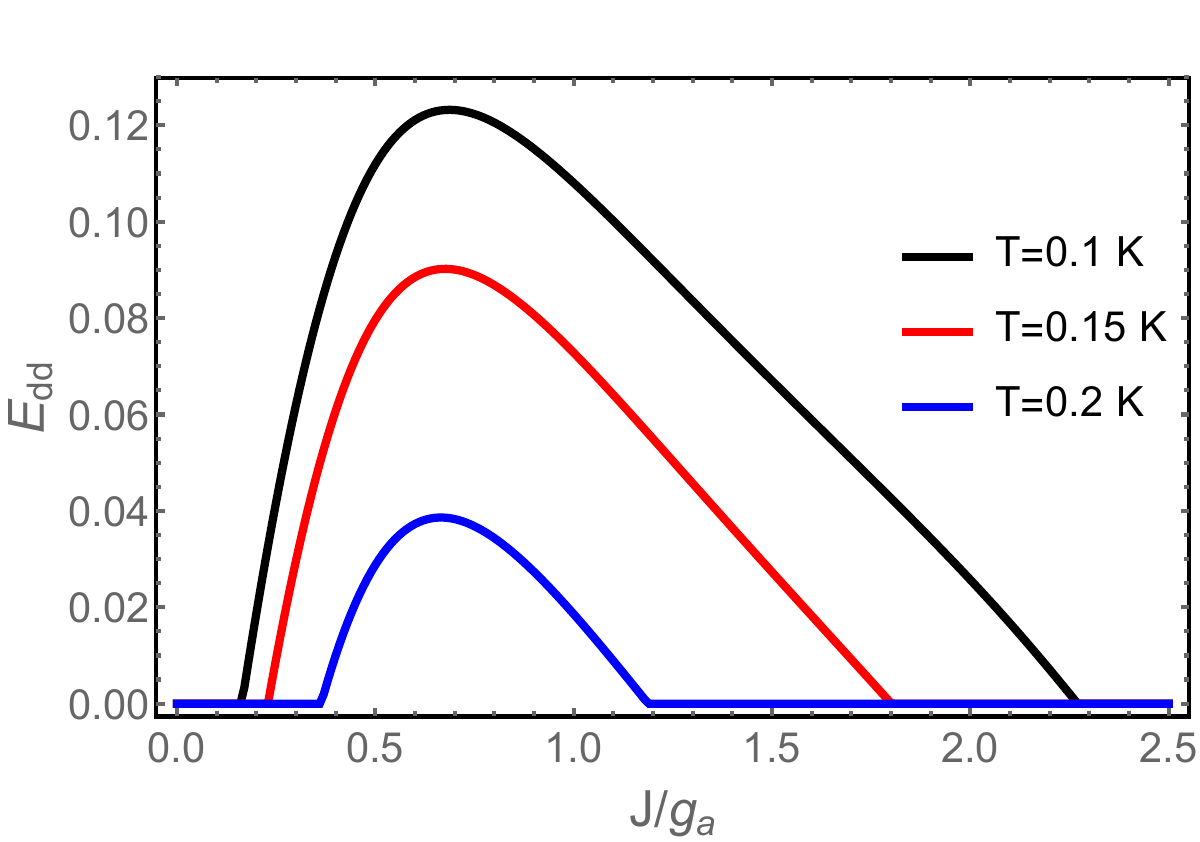}
\caption{Plot of the logarithmic negativity $E_{dd}$ between the two magnon modes versus the photon tunneling rate $J$ for differents values of the temperature $T$ with $\Delta_a=\omega_b$, $\Delta_d=0.4\omega_b$ and $\lambda=0.05g_a$.}
\label{fig4}
\end{figure*}

In Fig. \ref{fig4}, we plot the entanglement $E_{dd}$ of two magnon modes as a function of the photon tunneling rate $J$. This figure shows that the generation of the entanglement between the two magnon modes requires a minimum value of $J > J_{min}$. This can be explain by the entanglement Sudden birth \cite{ZFicek2006}. The entanglement is sensitive to the temperature $T$ as depicted in Fig. \ref{fig4}. We remark, the entanglement increases monotonically with increasing the photon tunneling rate $J$. Moreover, it once it achieves its maximum value star to decreases quickly with $J$. This degradation of the entanglement can explain by when the photon tunneling rate $J$ exceeds its optimal value induces decoherening effects on the magnon and becomes a significant effective thermal bath for the mechanical mode. In fact, increasing the photon tunneling is responsible of more thermal effects which induce the degradation of the quantum correlations between the two magnon modes.

\begin{figure*}[htb]
\includegraphics[width=\textwidth]{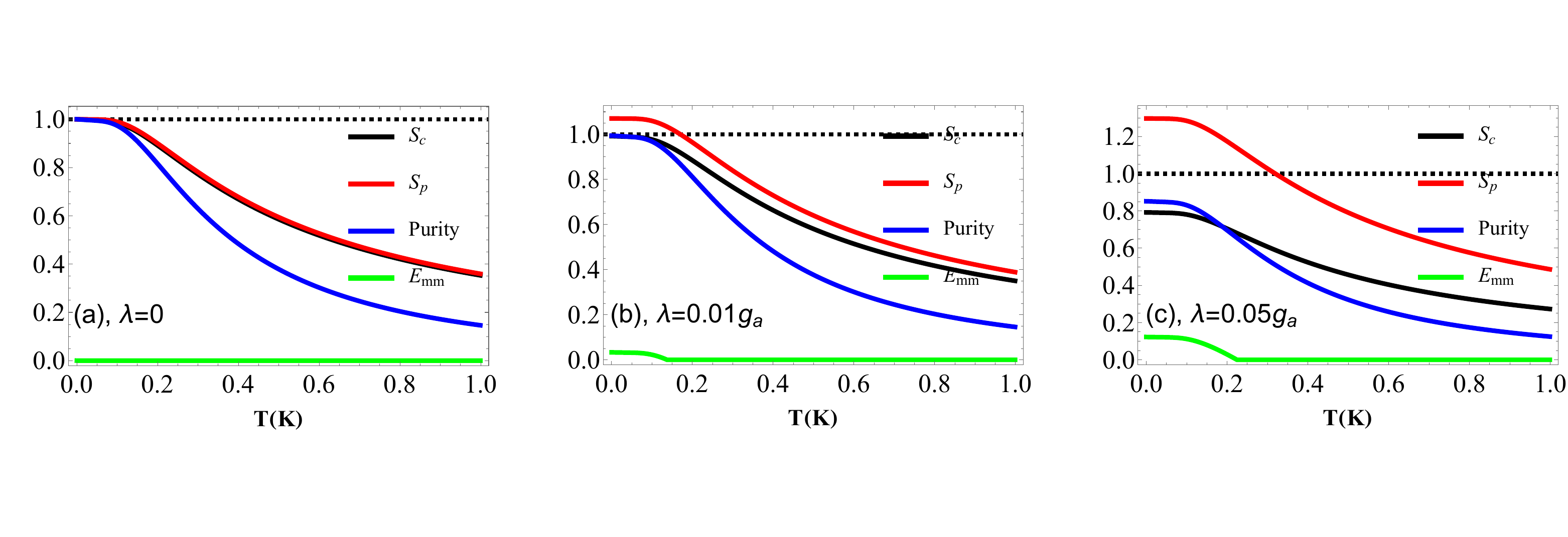}
\caption{Plots of the quantum complete synchronization $S_c$, quantum phase synchronization $S_p$, purity $P$ and negativity logarithmic $E_{dd}$ as a function of the temperature $T$ for various value of $\lambda$ with $\Delta_a=\omega_b$, $\Delta_d=0.4\omega_b$ and $J=0.5g_a$. }
\label{fig5}
\end{figure*}
We plot in Fig. \ref{fig5}, quantum complete synchronization $S_c$, quantum phase synchronization $S_p$, purity $P$ and negativity logarithmic $E_{dd}$ as a function of the temperature $T$ for various value of $\lambda$. Firstly, we remark that $S_c$, $S_p$, $P$ and $E_{mm}$ decrease quickly with increasing $T$ (decoherence phenomenon). By comparing curves of $E_{dd}$ with ones of $S_p$, we observe that the curves of $E_{dd}$ are very similar to that of quantum
phase synchronization $S_p$, especially when the parameter $\lambda$ is large enough and the temperature $T$ is very small. When $\lambda 0.05g_a$ the quantum phase synchronization is $S_p>1$ when $T<0.3$ K, this means that the fluctuation is squeezed between the two magnon modes. This can be from the enhancement of the nonlinearity of the system. In this case the entanglement the two magnon modes is enhances. Moreover, when $T>0.3$ K, the quantum phase synchronization is achieved ($0 < S_p < 1$) and the two magnon are separable ($E_{dd}=0$). Also, the curves of purity are very similar to that of quantum complete synchronization $S_c$ around cryogenic temperature. The purity is inversely correlated to $\lambda$, for example: when $\lambda=0$ the purity $P=1$ (the system of two magnon is pure), and when $\lambda = 0.05g_a$ the $0 < P <1$ (the system of two magnon became mixte). Furthermore, quantum complete synchronization $S_c$ equal to 1 for a very small range of temperature, i.e. the two magnon are fully synchronized. The synchronization $S_c$, $S_p$, and purity persist under thermal effects rather than entanglement, as depicted in Fig. \ref{fig5}.
\begin{figure*}[!htb]
\includegraphics[width=0.5\linewidth]{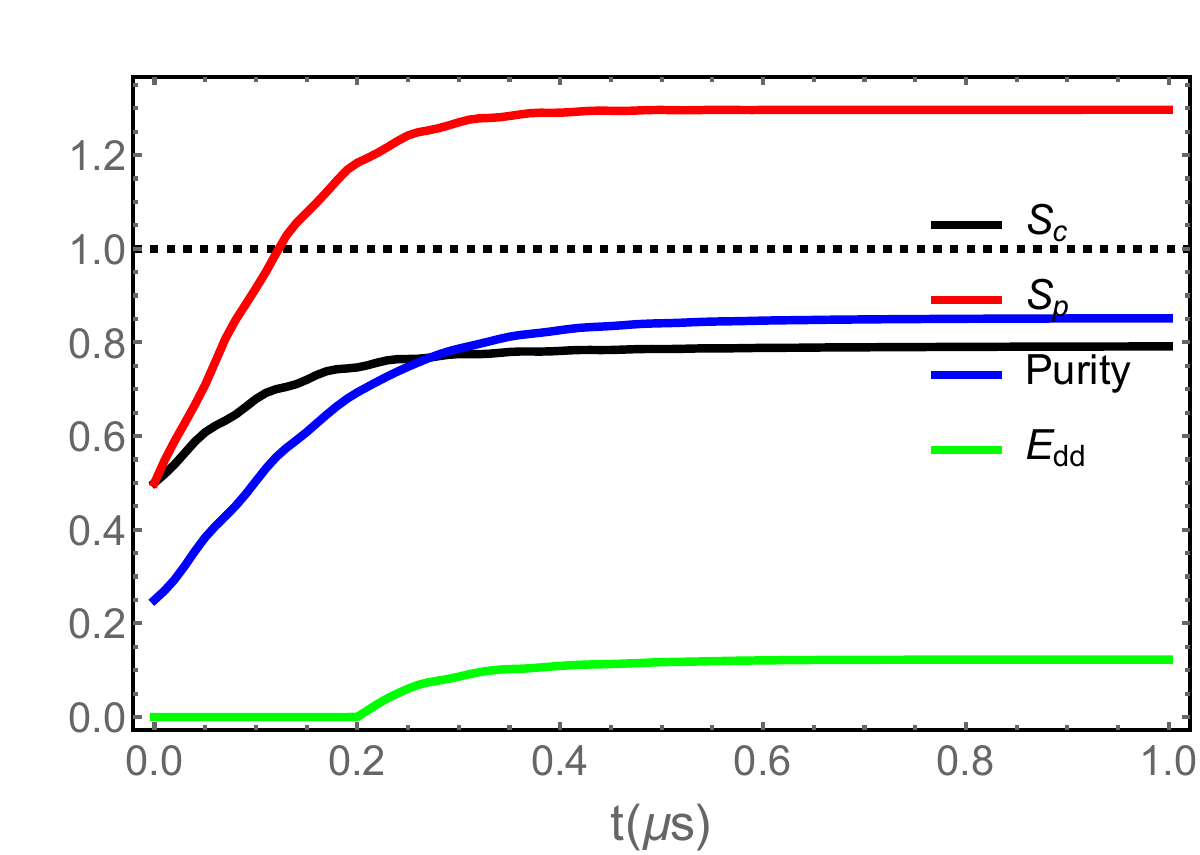}
\caption{Time evolution of the quantum complete synchronization $S_c$, quantum phase synchronization $S_p$, purity $P$ and negativity logarithmic $E_{dd}$ with $T = 0.1$ mK, $\Delta_a=\omega_b$, $\Delta_d=0.4\omega_b$, $J=0.5g_a$ and $\lambda = 0.05g_a$.}
\label{fig6}
\end{figure*}

In Fig. \ref{fig6}, we plotted the time evolution of quantum complete synchronization $S_c$, quantum phase synchronization $S_p$, purity $P$, and negativity logarithmic $E_{dd}$ between the magnon modes. We note that the quantum state of the two magnon modes is separable (zero entanglement) when $t<0.2\mu$s, even if quantum complete synchronization $S_c$ and quantum phase synchronization $S_p$ are achieved. By comparing the curves of quantum complete synchronization $S_c$, quantum phase synchronization $S_p$, and negativity logarithmic $E_{dd}$ and purity, we observe that the curves are very similar when $t>0.2\mu$s. Also, all measures ($S_c$, $S_P$ and $E_{dd}$) and purity $P$ tend to achieve their maximum value, i.e., steady state. The fluctuation between the two magnon modes became squeezed when $t>0.12\mu$s. Then, although complete or phase synchronization can exist without entanglement, it is certain that quantum synchronization is associated with entanglement.

\section{Conclusions} \label{Conc}

In summary, we have theoretically studied the enhancement of magnon-magnon entanglement in a cavity magnomechanical system through the squeezing of the magnon mode. We have thoroughly investigated the relationship between quantum synchronization, purity and quantum correlations in a dissipative magnomechanical system. Our results demonstrate a significant improvement in the entangled state of two magnon modes in two massive YIG spheres through magnon squeezing. Magnon squeezing leads to enhanced nonlinearity in the system, resulting in stronger non-classical correlations, particularly at low temperatures. We also show that thermal effects do not quickly weaken the entanglement between the two indirectly coupled magnon modes in the presence of magnon squeezing. Additionally, photon tunneling strengthens the shared quantum correlations between the two separated magnon modes. We have shown that quantum synchronization and purity are more robust against thermal effects than entanglement. Furthermore, the entanglement starts to increase when the purity is lower than 1 and when fluctuations are squeezed between the two magnons. We illustrate that the impact of $\lambda$ on achieving quantum complete synchronization exhibits comparable behavior to its effect on purity. However, $\lambda$ effect on quantum phase synchronization is more akin to its influence on quantum entanglement, as depicted in Figure (\ref{fig5}). Besides, we have seen that quantum synchronization and mixedness are associated with entanglement in both steady and dynamical regimes.

\end{document}